\documentclass[fleqn,10pt]{wlscirep}
\usepackage[utf8]{inputenc}
\usepackage[T1]{fontenc}
\usepackage{rotating}
\usepackage{float}
\title{Probabilistic Mixture Model-Based Spectral Unmixing}

\author[1,*]{Oliver Hoidn}
\author[1]{Aashwin Ananda Mishra}
\author[1]{Apurva Mehta}
\affil[1]{SLAC National Accelerator Laboratory, Menlo Park, CA-94025, USA}

\affil[*]{ohoidn@slac.stanford.edu}

\begin{abstract}
Identifying pure components in mixtures is a common yet challenging problem. The associated unmixing process requires the pure components, also known as endmembers, to be sufficiently spectrally distinct. Even with this requirement met, extracting the endmembers from a single mixture is impossible; an ensemble of mixtures with sufficient diversity is needed. Several spectral unmixing approaches have been proposed, many of which are connected to hyperspectral imaging. However, most of them assume highly diverse collections of mixtures and extremely low-loss spectroscopic measurements. Additionally, non-Bayesian frameworks do not incorporate the uncertainty inherent in unmixing. We propose a probabilistic inference approach that explicitly incorporates noise and uncertainty, enabling us to unmix endmembers in collections of mixtures with limited diversity. We use a Bayesian mixture model to jointly extract endmember spectra and mixing parameters while explicitly modeling observation noise and the resulting inference uncertainties. We obtain approximate distributions over endmember coordinates for each set of observed spectra while remaining robust to inference biases from the lack of pure observations and presence of non-isotropic Gaussian noise. Access to reliable uncertainties on the unmixing solutions would enable robust solutions as well as informed decision making. 
\end{abstract}

\begin{document}

\flushbottom
\maketitle
Mixing is ubiquitous in nature, ranging from the slow mixing in the earth’s mantle to manufacturing processes involving blending of viscous polymers or even the dispersion of solids into rubber\cite{ottino1989kinematics}. The process and mathematics of mixing are of great interest in physics and engineering applications, from mixing in process engineering to the chaotic mixing of horseshoe maps. The reverse of this process, referred to as unmixing, segregation or separation, is also of great interest but challenging. It involves recovering information lost through the changes in entropy due to mixing. The process is possible computationally if the mixing is incomplete (i.e., there is variation in the mixing fractions at different sites in the material), and sufficient spectral signal is visible from the individual mixture components. It is attempted almost entirely for cases where the equations governing mixing are linear\cite{heller1960unmixing}. Computational unmixing is critical for understanding and controlling processes like electrolysis refining\cite{zhang2012removal}, petroleum refining\cite{gary2007petroleum}, chemical analysis via chromatography\cite{smith2013chromatography}, besides others. 

Spectral unmixing techniques, which include but are not limited to Hyperspectral Imaging (HSI) \cite{chang2003hyperspectral, adao2017hyperspectral}, aim to identify and quantify material components in mixed signals by comparing sets of spectral measurements to a library of known signatures. Ideally, each pure component in a mixture must have a unique spectral signature. However, in many cases the conditions of the data are spatially or temporally-dependent, causing inherent variation in the endmember signatures. Consequently, the assignment of a unique spectral signature to each endmember is a simplification that may lead to limitations in the accuracy of the unmixing estimates. Techniques have been developed to handle the more general case, where endmembers are represented by a set of spectra \cite{bateson2000endmember, somers2012automated}.

Henceforth we will call the unique spectral signature of a component an `endmember'\cite{kale2019hyperspectral}. Each observation in such a set of measurements contains a `spectral footprint' corresponding to the underlying material mixture. The subsequent task is inverting this data to identify each endmember's unique spectral signature, as well as to quantify the proportion of each endmember in each mixture. This process, called 'unmixing', finds applications in diverse fields, from geology\cite{zhizhong2012review} and agriculture\cite{lu2020recent} to astronomy\cite{hege2004hyperspectral} and material sciences\cite{dong2019review}. In any context unmixing represents an ill-posed problem and the solutions may not be unique. In this light, it is essential to be able to represent the uncertainties associated with the unmixing process. The success of spectral unmixing relies on the selection of a spectral tool to analyze the data, the spectral distinctiveness of the endmembers, and the diversity of available mixtures\cite{keshava2002spectral}. 

Given a particular choice of spectral measurement tool, the most salient considerations are \emph{noise} and \emph{mixture diversity}. Additional challenges arise from factors such as non-linear mixing and data distortions. Noise, or more accurately noise relative to the spectral separation between endmembers, determines the ease of distinguishing hypothetical pure spectral signatures. Thus noisy observations can complicate extraction -- as can homogeneous mixtures, in which endmember extraction becomes nearly impossible even in the absence of noise \cite{chang2013hyperspectral}. Conversely, if endmembers are immiscible -- meaning the mixing fraction of an endmember in a given observation is either 0 or 1 -- extraction is trivial as long as the noise level is small compared to the spectral separation between endmembers. Here, we focus primarily on the challenging, but tractable, intermediate case of heterogeneously mixed linear observations.

To illustrate the property of mixture diversity, consider a color photograph represented as a set of multi-dimensional arrays of weights, with endmembers identifiable as distinct colors: red, blue, and green. A color photograph is an example of a heterogeneously mixed set of observations that can be mapped to points on a 3-simplex, and the maximally diverse color photograph is that which contains at least one each of a pure (or completely) red pixel, a pure blue pixel, and a pure green pixel (i.e., the vertices of the simplex).

The concept of an endmember extends beyond discrete colors in many scientific applications. The array of weights can be multi-dimensional, including non-spatial components like time, scattering vectors, chemical compositions, red-shifts, and more. Rather than viewing HSI as the sole technique, we adopt a broader perspective where the observations need not be a specific type of spatial map. Instead, they can be treated as an unordered set of spectra in which other descriptors (e.g., spatial parameters) are not necessarily needed for the endmember extraction, and thus the problem we are addressing is the more general task of spectral unmixing.

The analogy of a color photograph can be generalized, for instance, to an ensemble of partially mixed chemical species. Here, each observation of a set of spectral mixtures is a spectrum comprising discrete samples along spectral dimension $\lambda$, and each individual spectral signature corresponds to, for example, the electromagnetic spectrum from a different chemical species. An endmember comprises a singular spectral signature and its associated species. The goal of the analysis is to produce weights over the endmembers that are present in each observation, along with an approximation of each endmember's signature.

In the broad landscape of spectral unmixing, several techniques -- including geometric/simplex methods (e.g. N-FINDER, the most prevalent instance of these) and subspace reduction (e.g. nonnegative matrix factorization) -- have emerged as the dominant approaches to endmember extraction \cite{nfindr, miao2007endmember}. While it would be exhaustive to provide a comprehensive review of existing endmember extraction methods (for that, refer to Bioucas-Dias et al. \cite{dias}), our focus here is on comparing the popular simplex approach to a Bayesian model. The intent is to highlight the advantage of probabilistic modeling over point-estimate approaches. The Bayesian approach explicitly acknowledges and incorporates the epistemic uncertainties inherent to the unmixing process. Additional, Bayesian approaches can incorporate aleatoric uncertainties in the problem, arising for instance due to noise in the measurements. The final solutions engendered by such a Bayesian approach reflect the model's confidence in its prediction. Having access to such predictive uncertainties can assist in informed decision making under uncertainty.

\section*{\label{sec:methods}Methods}
\begin{figure}[H]
\centering
\includegraphics[width=0.5\textwidth]{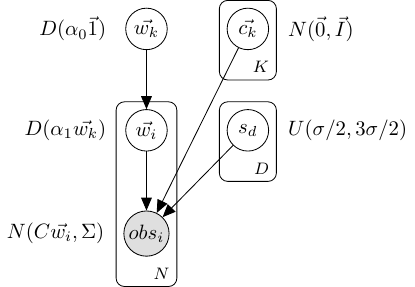}
\caption{Plate diagram for the Bayesian endmember mixture model. The top-level variables $\vec{w}_k$, $s_d$ and $\vec{c}_k$ are given by equations (\ref{eq:1}) and represent the prevalence of each mixture component, the extent of noise, and the endmember coordinates, respectively. `obs', for `observation', represents an observed mixture, of which $N$ samples from the underlying distribution constitutes one generated dataset.}
\label{fig:plate}
\end{figure}

\subsection*{Mixing Model}
Our approach begins with a simple physical model for mixing and observation. Suppose that each $k$th endmember has a global relative concentration $w_k$ (with $\Sigma_k w_k = 1; w_k > 0$) and that this global concentration is \emph{a priori} unknown. Given particular values of $w_1, w_2, ... w_K$ for the $K$ endmembers (noted $\vec{w}$) the mixing process combines the endmembers into a sample with endmember fractions $\vec{w}_i$, with the single constraint that the long-run average of $\vec{w}_i$ over $i = 1, 2, 3, ...$ is equal to $w_k$. 

The observation step of the model identifies each $\vec{w}_i$ with a multidimensional Gaussian distribution in a $D$-dimensional measurement space. A \emph{single} observation of a mixture is a point in this measurement space drawn from the corresponding distribution. 

To formulate this mixing process as a deterministic model, we would omit the top level in this implied sampling hierarchy (i.e. the generation of endmember frequencies from a global prior for relative concentrations) and interpret the endmember weights ($\vec{w_k}$) and coordinates ($\vec{c_k}$) as fixed, unknown parameters to be estimated via optimization with respect to a dataset and likelihood function.

\subsection*{Bayesian formulation}
Instead of developing a deterministic approach we formulate a hierarchical probabilistic model that will allow us to infer the mixture parameters using Bayesian methods. The structure of this model is given by the plate diagram in Fig. \ref{fig:plate}. 

The most salient considerations in the unmixing process are mixture diversity and noise. In keeping with this, at a high level, our model structure encodes the diversity of mixtures using a variable $\alpha$ that parameterizes distributions over component fractions in the data-generating process. Similarly, it captures the degree of noise with a second top-level parameter $\sigma$. 

Sampling is performed once per mixture component $k$ to determine endmember parameters (for a total of $K$, the number of endmembers); per $d$th spectral dimension to scale the variance of observation noise in each dimension (for a total of $D = K - 1$, the number of spectral dimensions); and per $i$th observed data point. The choice of $D = K - 1$ follows from the desire for a direct comparison between the Bayesian hierarchical model and N-FINDR. N-FINDR, in turn, specifies this value of $D$ because its convex set geometry formulation establishes a correspondence between the spectral space and the rank $K - 1$ space of allowed mixing coefficients.

This coupling of $D$ to $K$ simplifies the approach but would, in a practical setting, require prior dimensionality reduction of the data. Second and more significantly, we have assumed that $K$ is known. The suitability of this assumption is likely domain-dependent. To take a particular example, in the context of phase identification in materials science the assumption may be reasonable to the extent that $K$ is constrained by considerations such as the Gibbs phase rule. The prior distributions for the variables $\vec{global\_weights}$, $\vec{locs_k}$ and $scale_d$ are 

\begin{equation}\label{eq:1}
\begin{split}
\vec{global\_weights} &= \vec{w_k}  \sim Dir(\alpha_0 \vec{1}),\\
\vec{locs_k} &= \vec{c_k}  \sim N(\vec{0}, \vec{I}),\\
scale_d &= s_d  \sim U(\sigma / 2, 3 \sigma / 2),
\end{split}
\end{equation}
where $Dir(\alpha_0 \vec{1})$ is a symmetric $K$-dimensional Dirichlet distribution over global endmember frequencies; $N$ is a $D$-dimensional Gaussian over endmember coordinates; and $U$ is a uniform distribution used to rescale samples from an LKJ distribution over observation noise correlations matrices. Here, the order $K$ of the Dirichlet distribution is equal to the number of endmembers (mathematically, the number of categories). For a single sample, the underlying distribution over phase fractions is a Dirichlet distribution that depends on both the global weights and a second concentration hyperparameter, $\alpha_1$, a measure of the degree of mixing:
\begin{equation}
phaseweight_i = \vec{w_i}  \sim Dir(\alpha_1 \vec{w_k})\
\end{equation}

The use of two Dirichlet distributions in the hierarchical model -- one for the global endmember frequencies and one for the density of endmember mixing fractions within a particular mixture -- follows from non-negativity and sum-to-one constraints applied to the endmember global weights and mixture fractions, respectively.

Finally, the observation likelihood is a multivariate Gaussian with mean vector composed of the matrix product between the $D \times K$ endmember coordinates and the local phase weight vector $\vec{w_i}$, and covariance given by $\Sigma{(\vec{s})}$ (itself dependent on both the LKJ prior and the scale parameters $\vec{s} = (s_0,... s_d)$):
\begin{equation} \label{likelihood}
obs_i = y_i = N(C \vec{w_{i}}, \Sigma),
\end{equation}
where $C$ is the $D \times K$ matrix containing the endmember coordinate vector $\vec{c_k}$ for all values of $k$.
A single simulated dataset is produced by first drawing $\vec{c}$, $\vec{s}$ and $\vec{w}$ from the respective Bayesian priors and then generating $N$ samples from the likelihood, eq. \ref{likelihood}. 

In a bird's eye overview of our method, we generate simulated data of multiple points and then rely on two methods for inference of end-members, Hamiltonian Monte Carlo (HMC) and variational inference (VI). Once a probabilistic model has been derived and established in the form of the model expression, one needs to infer the distributions of the parameters in the model expression. This inference is generally intractable for any realistic model due to the integrals involved in the Bayesian estimation\cite{murphy2012machine}. To circumvent this intractability alternative approaches are utilized. The gold-standard in deriving parameter uncertainty have been sampling based Markov Chain Monte Carlo (MCMC)\cite{murphy2022probabilistic} methods, such as Metropolis Hastings. However, these have poor scaling and are computationally infeasible for high dimensional problems. Recent developments of Hamiltonian Monte Carlo and specifically, the No U-Turn Sampler (NUTS), have transformed this status quo \cite{betancourt2017conceptual}. HMC avoids the random walk dynamics of classical MCMC algorithms, by using the information in the gradient of the target distribution to inform sampling. NUTS regulates the frequency of sampling and together, HMC-NUTS enables sampling based uncertainty estimation approaches to be extended to higher dimensions. However, in many cases with higher dimensions, HMC-NUTS may still not scale well and be time consuming. Thus, we also test inference using Variational Inference using the mean field approximation. Stochastic Variational Inference\cite{hoffman2013stochastic} reduces the sampling problem to an optimization exercise wherein the divergence between the posterior distribution and a family of parameterized distributions is minimized. This enables the use of classical gradient based optimizers to solve the problem rapidly and scale to even higher dimensions. However, in line with the no free lunch theorem, these advantages in scaling are accrued at the cost of being unbiased. Based on the assumptions made during VI to simplify the problem for computational and scaling expense, the final solution can be biased. This may lead to over-confidence and under-prediction of the uncertainty. To this end, we compare and contrast the solutions that we generate via HMC-NUTS and mean-field VI.

\subsection*{Simplex formulation}
Here, we compare and contrast the Bayesian approach introduced in this section to a traditional approach to unmixing. In the simplex formulation and under the assumption of linear mixing of spectral signatures, the number of sample points in the spectral dimension $\lambda$ as $D$, and the presence of up to $K$ endmembers, the fractional abundances map to points in the $K$-simplex. Furthermore, the $K$ endmembers define a $D$-volume (in a $D-$ dimensional spectral space) that fully encapsulates all observed spectra. If we assume noise-free conditions and the presence of pure observations in the dataset for each endmember, where the fractional abundance of that endmember is $1$, it follows that the endmembers correspond to a choice of $K$ observations from the dataset that define the maximal volume. This maximal volume condition is the premise behind the N-FINDER algorithm and other simplex methods.

While the above simplified model offers a clear interpretation of the mixing process, its application in endmember extraction necessitates two assumptions: that observed spectra are free of noise and that pure endmembers are present in the data. While some previous methodologies relax one or both of these assumptions, they typically do not factor in uncertainties related to inferred endmember parameters or mixing weights. An instance of such an approach is the noise-adjusted principal components (NAPC) methodology \cite{napc}.

In contrast, our proposition involves a mixture model for probabilistic estimation of endmember parameters that includes (Gaussian) observation noise and incomplete sampling of the endmember simplex (i.e., reduced diversity). Additionally, this probabilistic framework incorporates the epistemic uncertainty inherent to the unmixing problem due to its ill-posedness. Using this model and simulated observations, we approximate a full posterior distribution over endmember spectra via Hamiltonian Monte Carlo and Variational Inference. Upon comparing endmember reconstruction using inference under our model to N-FINDER -- a representative example of a deterministic endmember extraction method -- we find the Bayesian approach to be robust to noise and incomplete observation of pure samples. As a result, our model offers significantly more accurate endmember extraction, as quantified by the reconstruction error of endmember coordinates.

\begin{sidewaysfigure}
    \centering
    \includegraphics[trim={2.5cm 0 2.5cm 0},clip, width = 1.\textwidth]{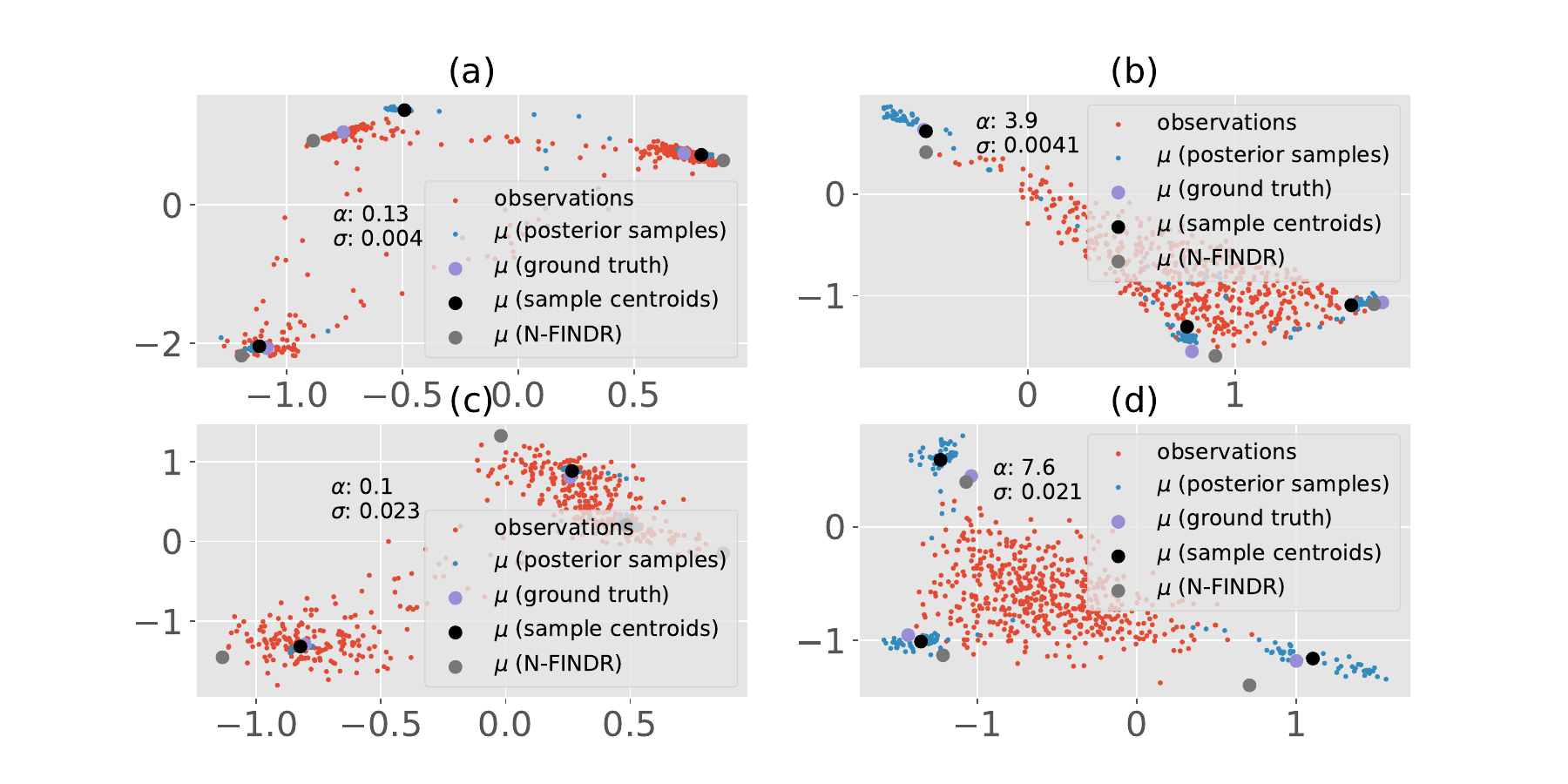}
    \caption{Comparison of endmember coordinates extracted using N-FINDER vs. through sampling of the model posterior using HMC and the no-U turn (NUTS) sampler in Pyro \cite{pyro}. The panels (a) and (b) on the top row represent the results of endmember extraction on datasets generated under different conditions of the concentration hyperparameter $\alpha_1$: small and large respectively, coupled with a small noise scale parameter ($\sigma$). The bottom row panels (c) and (d) depict inference results for large noise scale parameter ($\sigma$), again contrasting between small and large $\alpha_1$.}
    \label{fig:comparison}
\end{sidewaysfigure}

\section*{Results}
\begin{sidewaysfigure}
    \centering
    \includegraphics[trim={7cm 0 0 0},clip, width=1.1\textwidth]{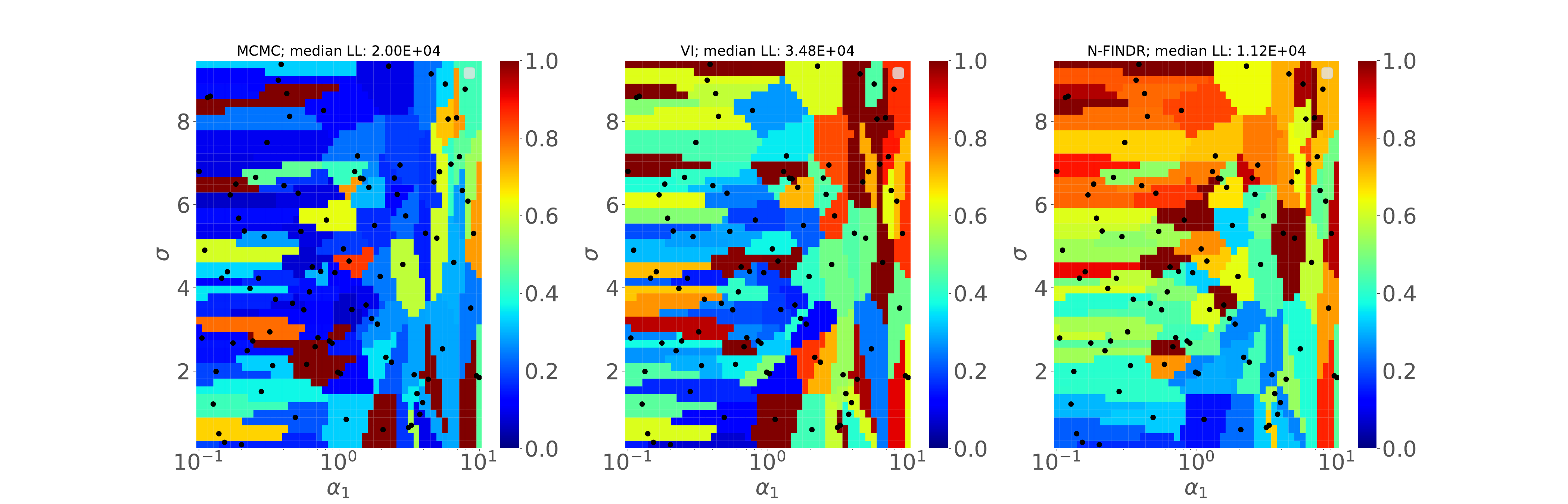}
    \caption{L2 reconstruction error for endmember coordinates via HMC posterior samples (left), mean field variational inference (center),  compared to N-FINDER (right), with varying values of the hyperparameters $\alpha_1$ and $\sigma$. The reconstruction error is a sum in quadrature of the L2 distance between }
    \label{fig:heatmap}
\end{sidewaysfigure}

\begin{sidewaysfigure}
    \centering
    \includegraphics[trim={1.55cm 0 0 0},clip, width=1.1\textwidth]{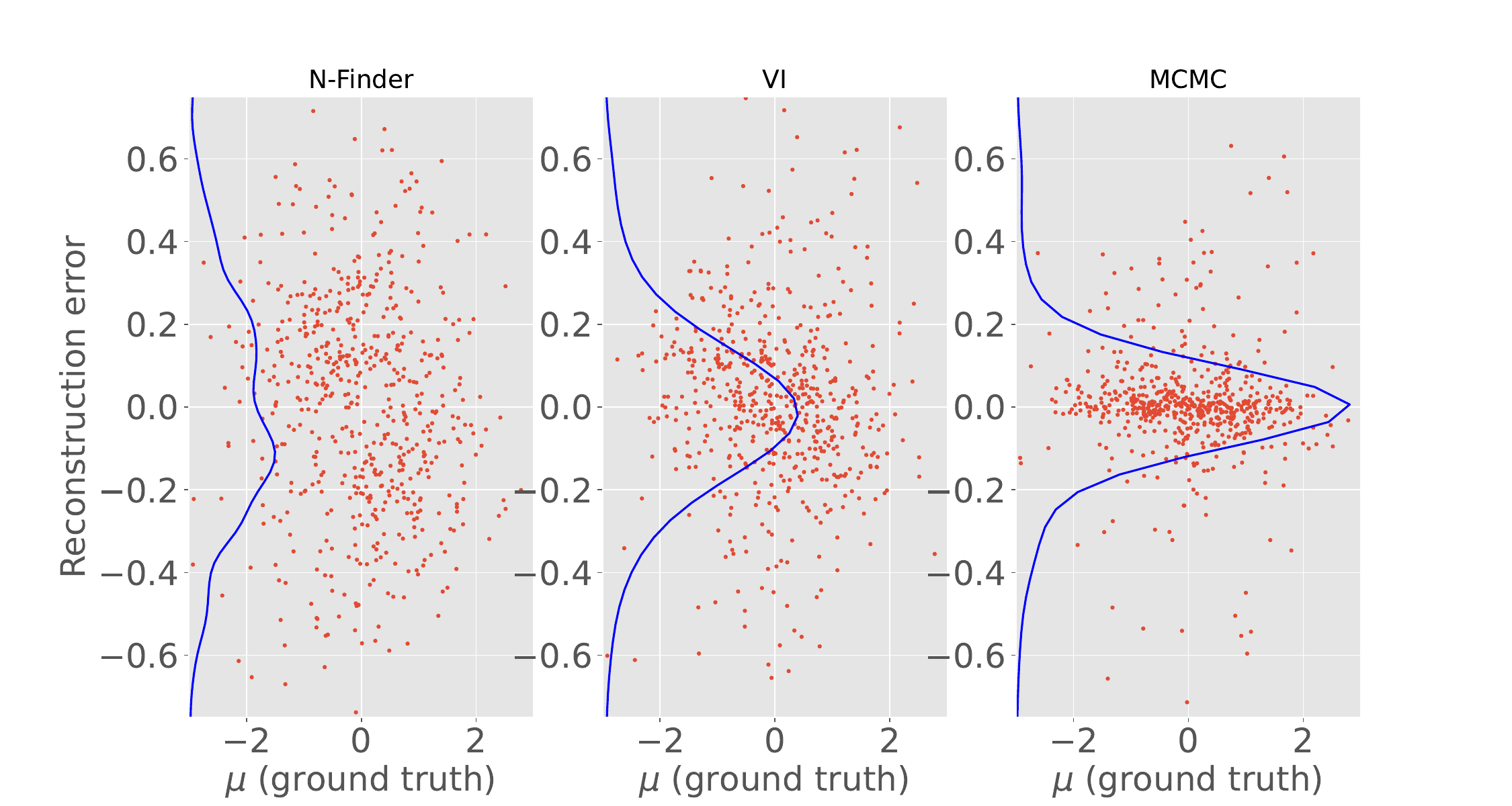}
    \caption{Scatter plots of reconstruction error (horizontal axis) and ground truth (vertical axis), concatenated over the individual scalar components of the endmember coordinates under HMC sampling (right), mean field variational inference (center), and N-FINDER (left). The blue curve in each panel is a smoothed density estimate along the horizonal axis -- it represents the empirical probability distribution, over all datasets, for the scalar difference between the endmember coordinate estimates and ground truth.}
    \label{fig:residuals}
\end{sidewaysfigure}



After generating $100$ simulated datasets of $N = 500$ observations each in the above manner, we infer end-member coordinates in two ways: first, by using the model distribution function to sample from the posterior over model parameters with the no- U turn (NUTS) Hamiltonian Monte Carlo sampler; second, with mean-field stochastic variational inference (SVI) using the ADAM optimizer and a learning rate of 0.005. To mitigate failures to converge (a common issue with mixture models) we run multiple rounds of inference with different random initializations. We select between 3 starts per dataset for HMC and 10 for VI; in each case, we retain only the inference round that produces the highest log likelihood for the dataset under the posterior samples (HMC) or approximation (VI). Per dataset, the sampling and inference ran for a total of 10 minutes on 12 CPU cores.

Figure \ref{fig:comparison} juxtaposes the endmember coordinates determined through N-FINDER and the hierarchical Bayesian model using HMC with the NUTS sampler. Specifically, it shows representative outcomes of these two methods under the four possible extreme conditions of hyperparameters $\alpha_1$ and $\sigma$: low $\alpha_1$ with low $\sigma$, low $\alpha_1$ with high $\sigma$, high $\alpha_1$ with low $\sigma$, and high $\alpha_1$ with high $\sigma$. This comparison provides a qualitative contrast of the data distribution and the respective behaviors of the two endmember extraction approaches in different corners of the space of parameters $\alpha_1$ and $\sigma$, which, we reiterate, capture the degree of mixing and severity of noise in the formulation. In this comparison N-FINDER serves not as an a example of a state of the art model but rather as a representative instance of deterministic end-member extraction. 

Further, Figure \ref{fig:heatmap} compares the reconstruction accuracy for 100 instances of this simulation-inference procedure, with 100 distinct combinations of the hyperparameters $\alpha_1$ and $\sigma$. For VI and HMC we use monte carlo samples from the posterior to estimate the central tendency (ie, estimated mean from MC samples) for the prediction. This is used as the approximation for the point prediction for the Bayesian approaches. With N finder we use monte carlo samples over different seeds, and average these samples for the mean N-Finder prediction. In each of the cases, due diligence was carried out to ensure that the number of MC samples was sufficient for a robust final prediction.Subsequently, we use the Frobenius norm of the difference between the point prediction and ground truth as our figure of merit for the reconstruction accuracy.

Given the above centroid-distance figure of merit, \ref{fig:heatmap} compares the reconstruction accuracy of HMC-NUTS (right) and mean field variational inference (center) to that of the N-FINDER baseline (right). High values of the concentration parameter $\alpha_1$ correspond to low mixture diversity -- that is incomplete sampling by the data of the endmember simplex. The combination of high noise ($\sigma$) and high $\alpha_1$ is the most difficult case for inference and consequently results in poor reconstruction accuracy for all approaches. Conversely, the methods exhibit better performance when $\sigma$ and $\alpha_1$ are both small. In regions where either of $\sigma$ or $\alpha_1$ is large, however, HMC posterior sampling performs much better than N-FINDER. The improved reconstruction when pure observations are missing (high $\alpha_1$) or the noise amplitude is high (large $\sigma$) conforms to expectations of the conditions under which one should expect benefits from the mixture model approach. 

Finally, we investigate posterior approximation with variational inference as a computationally-cheaper alternative to HMC. Figure \ref{fig:residuals} presents a three-way comparison of the reconstruction error for end-member coordinates under HMC, VI and N-FINDER. As is evident in the figures, HMC and N-FINDER generate the highest and lowest quality reconstructions, respectively. Further, mean field VI leads to reconstructions that lie between HMC and N-FINDER. However, VI provides an approximately 100-fold advantage in computation time over HMC, on average. 

\section*{Summary \& Future Directions}
In summary, we applied a probabilistic mixture model for the extraction of endmembers from spectral mixtures. Our approach formulates the retrieval of endmember spectra and per-observation endmember weights as a Bayesian inference problem using relatively uninformative priors for the phase-mixing and observation processes. Using simulated datasets we have sampled the model posterior with HMC and found that the resulting recovered endmember parameters are more robust to observation noise and absence, in the data, of pure endmember spectra compared to the popular N-FINDER algorithm. This Bayesian approach provides a principled framework for incorporating relevant prior information without introducing undue assumptions and opens the door to other intrinsically probabilistic analyses, such as uncertainty quantification. 

Both the baseline model (N-FINDER) and our Bayesian approach assume that the number of endmembers is known. Unlike deterministic approaches such as N-FINDER, the probabilistic formulation would allow one to relax this potentially-onerous assumption by replacing the fixed number of mixture components and corresponding endmember weights by a set of samples from a stick-breaking distribution\cite{teh2010dirichlet}, as in a Dirichlet process mixture model. Assumptions about $K$ would thus be condensed into a single hyperparameter, the concentration parameter of the stick-breaking process. Additionally, our approach can be generalized to incorporate more flexible, non-linear mappings using Bayesian Neural Networks (BNNs) to infer embeddings.


\bibliography{sample}

\section*{Acknowledgements}
This work was performed and partially supported by the US Department of Energy (DOE), Office of Science, Office of Basic Energy Sciences Data, Artificial Intelligence and Machine Learning at the DOE Scientific User Facilities program under the MLExchange Project (award No. 107514). Aashwin Ananda Mishra was partially supported by the SLAC ML Initiative.

\section*{Author contributions statement}
O.H. and A.A.M. wrote the manuscript with input from A.M. All authors discussed results, edited the manuscript, and gave final approval for publication. 

\section*{Ethics declarations}
The Authors declare no Competing Financial or Non-Financial Interests.

\section*{Code and Data Availability Statement} 
The code and datasets generated during and/or analysed during the current study are available from the corresponding author on reasonable request. 

\end{document}